# Finite Information Numbers through the Inductive Combinatorial Hierarchy

Theophanes E. Raptis[abc]

[a]National Center for Science and Research "Demokritos", Division of Applied Technologies, Computational Applications Group, Athens, Greece.

[b]University of Athens, Department of Chemistry, Laboratory of Physical Chemistry, Athens, Greece.

[c]University of Peloponnese, Informatics and Telecommunications Dept., Tripolis, Greece.

**Abstract:** We report on a recent conjecture by Gisin on a restriction of physical processes in sets of finite information numbers (FIN) and further analyze the entropic constraint associated with the proposed algorithm. In the course, we provide a decomposition of binary entropies in a pair of fractal sequences as functional composites of binary digit-sum functions and we construct a unique formula and an abstract partition function for these. We also prove based on previously introduced tool of the inductive combinatorial hierarchies that the naturally inherited self-similarity of the resulting hierarchy of entropic sets contains equivalence classes providing unlimited symbolic series for satisfying the demand posed by the FIN conjecture.

In a recent original criticism by Gisin [1], an argument is made of a finite space volume being unable to hold an infinite amount of information based on previous considerations of the Bekenstein entropy being bounded. Because of the need to retain the previously established mathematical toolbox of the standard dynamical equations defined on a standard continuum, Gisin asks to consider only numbers of finite information content (FIN) of which an initial part of some bit length *L* is given by a determined probability initially set to 1 for a few bits, followed by a decrease of that probability while a residual infinity of bits becomes perfectly random and thus inconsequential for the dynamics apart from the fact that any truly chaotic dynamical system would appear perfectly randomized. The resulting numbers appear then to have finite entropy and to also have a closure property for any finite computation. To further shed light in the nature and the possible ramifications of Gisin's conjecture, we need to examine in detail certain global properties of the restricted subset of reals as offered by the original recipe in [1]. In what follows we examine first a decomposition of the generic binary entropy function as a composite over the binary digit-sum function $\Sigma_2(v)$ where *v* any positive integer and show by induction the existence of a unique resolvant formula. We do so by utilizing a previously introduced tool in the form of the Inductive Combinatorial Hierarchy as briefly presented in section 4 of [2].

Work on the digit-sum function has a long history as it permeates many important areas in the general theory of automata, algorithm analysis and coding theory.

Trollope [3] and Delange [4] studied the existence of a closed formula for successive summands of this function which will be important in what follows, based on a continuous, nowhere differentiable 1-periodic fractal function of which the Fourier expansion involves the Riemann Zeta function. Later, Flajolet *et al.*, further analyzed this function for higher radices (alphabet bases) using methods from analytic number theory via Mellin transforms [5], [6]. Gelfond [7] had already prescribed three basic problems related to the distribution of digit-sum functions also in relation with primes for the first of which Besineau offered an asymptotic solution at 1972 [8] till the final solution by Dong-Hyun Kim at 1999 [9] while the other two were only solved at 2009 [10] and 2010 [11] by Mauduit and Rivat which further improved on the use of Fourier based methods. Adam-Watters and Ruskey provided systematic methods for finding relevant generating functions [12]. Originally Bellman and Shapiro [13] and later, Mauduit and Sarkozy [14] as well as Madritch [15] studied properties of sets with analogous order. Kamiya and Murata [16] also generalized previous work by Flajolet and Ramshaw [17] showing that successive differences of $\Sigma_b$ are automatic. In a recent thesis by Müllner [18], several important connections between the third Gelfond problem, automaticity and deterministic dynamical systems are exposed, extending previous work by Drmota, Mauduit and Rivat.

In section 2, we utilize a particular redefinition of the proposed FIN structure via decomposition on a countable set and a residual so as to be able to produce global maps of entropic values on a hierarchy of integer intervals and their corresponding bit strings which leads to a redefinition of binary entropy in terms of certain direct functional composites over digit-sums. We also extract a form of abstract partition function from an exact formula for the digit-sum function. In section 2, we explore the scaling relationships of the resulting entropy sets and show that they have a self-affine character that allows their separation into equivalence classes which should be relevant for any entropy bound in the FIN conjecture. We conclude with a discussion of a possible filamentary topology that could occur if adopting FIN as a valid subset for physically arithmetizable structures.

## 2 The set of "Gisin reals"

Let the set of Gisin reals or FIN be $\Sigma$, satisfying $\Sigma \subset R$. One can always write all real numbers in the equivalent form

$$x \to x' = 2^k x = \nu + \varepsilon, \quad \nu \in [0,...,1) \subset Z, \quad \varepsilon = 0.\sigma_1\sigma_2... << 1 \in [0, 1-\varepsilon), \quad \sigma_i \in \{0,1\} \qquad (1)$$

We assume that every function can also be transformed as $f(x) \to f(2^{-k} x')$. As is evident, in this representation we still have a dense residual subset with same cardinality with $\varepsilon$ the radius of disks surrounding all members from a set of rationals. The restriction posed by the FIN recipe further imposes a "thinning" of the reals so as to restrict the values of $\nu, \varepsilon \in \Sigma$ so as to have a constant entropy value for all $\nu$ and also restrict the $\varepsilon$ values to maximal entropy. This necessarily removes some rationals and their disks creating "holes" on the real line. In higher dimensions, such a removal process would lead to a kind of disconnected topology and it might point towards multiply connected domains possibly with a noisy, multifractal structure like a finely thin network.

We now turn our attention to the countable set of $v$ values and use a direct mapping to an equivalent set of integers in the $[0,\ldots,2^k\text{-}1]$ via the polynomial representation for the inverse powers of *2* as symbols on a binary alphabet. To simplify the presentation we shall further restrict $v$ in subsets of $N$ leaving sign as a separate bit which can also be incorporated via a mapping as $(-)^{\mathrm{mod}(v^*,2)}$ where $v^*$ stands for the associated mirrored integer with all bits reflected. One immediately observes that for any $k$, the inclusive hierarchy of intervals $S_0 \subset S_1 \subset \ldots \subset S_k$ and its associated bit string dictionaries are closures for the totality of possible bit states of $v$ given some $k$, for which it is possible to fix certain measures. Most often, these turn out to be irregular fractal sequences in which case introducing the proposed varying probability for each bit as in [1], leads to a particular type of sieve for selecting bits.

The additional restriction required for the binary entropy of the integer part $v$ can be found via a global map over the interval $S_k:[0\ldots2^k\text{-}1]$. We shall prove that any such necessarily forms a fractal sequence over that interval. Specifically, we fix the binary probabilities as

$$p_1 = \Sigma_2(v)/k, \quad p_0 = 1 - p_1, \quad \forall k \in S_k \tag{2}$$

where $\Sigma_2$ stands for the number theoretic Digit-Sum function, a well known fractal sequence over any interval $S_k$ [], [] taking values inside $[0,\ldots,k]$. Using the complementary sequence $\Sigma_2^0(v) = k - \Sigma_2(v)$ and replacing into the standard Shannon formula results in the expression

$$H_k(v) = -\frac{\Sigma_2(v)}{k}\log_2\left(\frac{\Sigma_2(v)}{k}\right) + \frac{\Sigma_2(v)}{k}\log_2\left(\frac{\Sigma_2^0(v)}{k}\right) - \log_2\left(\frac{\Sigma_2^0(v)}{k}\right) \tag{3}$$

Expanding logarithms of ratios results in two equal and opposite terms $\mp\Sigma_2(v)\log_2(k)/k$ which cancel leaving after simplification the symmetric summand

$$H_k(v) = -\frac{1}{k}\{h(\Sigma_2(v)) + h(\Sigma_2^0(v))\} + \log_2(k) \tag{4}$$

where $h(x) = x\log(x)$. We notice that the non-monotonous increasing function $\Sigma_2$ always gets maximized at the end of each $S_k$ interval corresponding to an all ones pattern of a Mersenne integer $(2^k - 1)$. We also notice the internal symmetry of every such set due to the fact that any bitwise NOT complement $\bar{v}$ with opposite bits evidently satisfies $v + \bar{v} = 2^k - 1$ so that $\Sigma_2^0(v) = \Sigma_2(\bar{v}) = \Sigma_2(2^k - v - 1)$. This effectively makes $h(\Sigma_2^0(v))$ equivalent to a simple left to right flip of the graph of

$h(\Sigma_2(v))$ which gets maximized at the all zeros pattern. We may introduce a positive symmetric fractal sequence

$$H^0(v) = h(\Sigma_2(v)) + h(\Sigma_2^0(v)) = h(\Sigma_2(v)) + h(\Sigma_2(2^k - v - 1)) \qquad (5)$$

to represent the essential part of any such entropy as

$$H_k(v) = \frac{1}{k}\left(h(k) - H^0(v)\right) \qquad (6)$$

An example of this term is given in figure 1(a) for $k = 8$ and of $H^0(v)$ in 1(b) on the same interval. We see that there is a hierarchy of entropies for any $S_k$ which is solely a function of a fundamental fractal sequence. The asymptotic behavior of any such $H^0$ strictly depends on the asymptotic behavior of the function $\Sigma_2$. In what follows we shall provide an interesting case of an individually computable resolvant formula which can be found with the aid of the hierarchy and which to the author's knowledge does not appear in the OEIS database [19]. This can then be used to construct a hierarchy of partition functions over all string patterns.

The use of the hierarchical method requires finding by induction a uniformly computable recursive generator over any of the $S_k$ intervals and/or an individually computable resolvant for any sequence obtainable via some other bitwise computation. In full generality, we define a *parsing* map $\pi: S_k \times F \to S'$ where $S'$ is an arbitrary arithmetic interval as the codomain, and which acts on the associated bit string dictionary as a $k \times b^k$ array where each string is interpreted as an index running from 0 to $b$-1 with $b$ the alphabet base or radix and each index corresponding to a particular member of an ordered set of $b$ distinct functions from a functional lookup table $F$ (FLUT) so as to produce all possible functional compositions. We remind that such FLUTs can easily be build using anonymous functions or "*lambdas*" as they are called in the language of Church's calculus [20]. Blocks of same symbols obviously correspond to composition exponents. In the case of the digit-sum we just need a list like $\{f_0, g_1\}$ where $f_0$ is the identity and $g_1(x) = x + 1$. For any $S_k$ the associated $\Sigma_2$ is bounded in the $[0,\ldots,k]$ interval.

It is trivial to express the recursive form of $\Sigma_2$ since any pattern shall have only one new bit as 1 in every new exponential interval. In the language of [2], there is a single reproducing map as $K(\{x_i\}) = \{x_i\} + 1$ leading to $\Sigma_2(S_{k+1}) \leftarrow \{\Sigma_2(S_k), \Sigma_2(S_k)+1\}$ with $S_0 = \{0\}$, $\Sigma_2(S_0) = 0$ and the set union is taken as concatenation to keep the sequence ordered. In order to perform an effective interpolation with a unique formula for large intervals we have to apply a number of sequence foldings over successive intervals where periodicities appear until sufficiently simple relationships can be applied between previous and next columns. At a first level, repetition of 4 symbols across large chunks of this sequence reveals a method to imitate the basic pattern. It suffices to create an original sequence as $s_0(v) = \lfloor v/2 \rfloor + \mathrm{mod}(v,2)$ where $\lfloor \ \rfloor$ denotes the floor function. By folding this sequence via reshaping it into a 4 x $2^{k-2}$ matrix shows stable differences of value 2 over all columns for any interval. It suffices then to recursively subtract a number of periodic Boolean square functions $r_i(v, p_i)$ of increasing period as $2^i$ with $i = 2,...,k-2$ and coefficients as $2^i - 1$, resulting in the discrete integral form

$$\Sigma_2(v) = s_0(v) - \sum_{i=1}^{k-2}(2^i - 1)r_i(v, 2^i), \quad \forall v \in S_k \tag{7}$$

An analytical expression for the $r_i$ terms can be given with the aid of a general formula for 2-periodic Boolean sequences as $\theta(p_0 - \mathrm{mod}(v, p)), \ p > p_0$ where, $\theta(x)$ is the Heaviside step function and we have to choose periods as $p_0 = 2^i, p = 2p_0$. We also notice that the same functions can be used to construct all rows of any $k$ x $2^k$ array of deployed bit strings associated with any $S_k$ interval.

It is now possible to propose a kind of partition function given a selection of partial summands over exponential intervals as normalizing factors. We know that $\Sigma_2(2^k - 1) = k - 1$ and $\Sigma_2(2^k) = 1$ and we need to find the cumulants over any $S_k$. The expression given by Trollope and Delange has the explicit form

$$\sum_{v=0}^{n-1}\Sigma_2(v) \to (1/2)h(n) + nF_1(\log_2(n)) \tag{8}$$

where $F_0$ a continuous, periodic fractal function. Choosing $n = 2^k$ for each summand corresponding to an interval $S_k$ immediately leads to

$$\Sigma_{2^k}(S_k) = \sum_{v=0}^{2^k-1}\Sigma_2(v) \to k2^{k-1} + 2^k F_1(k) \tag{9}$$

In an interesting diversion from the original, recent work by Kruppel [] points towards a new expression for the fractal $F_1$ function in terms of the Takagi's [] continuous, non-differentiable fractal function as

$$F_1(k) = -k/2 - 2^{-(k+1)}T(2^k) \tag{10}$$

Replacing back in (8) results in

$$\sum_{v=0}^{2^k-1}\Sigma_2(v) \to k(2^{k-1} - 1/2) - (1/2)T(2^k) \tag{11}$$

The Takagi's function is a solution of the functional equations

$$T(x) = 2T(x/2) - x = 2T((x+1)/2) + x - 1 \tag{12}$$

The asymptotic limit of these solutions gives rise to the self-similar Blancmange curve, a member of the more general Takagi-Landsberg curves [22], [23], [24].

Armed with either (9) or (11) we can readily write for the abstract partition function in any $S_k$

$$Z_2(\nu) = \frac{\lfloor \nu/2 \rfloor + \mathrm{mod}(\nu, 2)}{\Sigma_{2^k}(S_k)} - \frac{1}{\Sigma_{2^k}(S_k)} \sum_{i=1}^{k-2} (2^i - 1) r_i(\nu, 2^i), \quad \forall \nu \in S_k \tag{13}$$

One can now derive various pseudo-thermodynamic quantities defined over the hierarchy of strings using (13) most of which will necessarily inherit the fractal nature of the defining functions albeit with occasional deformations caused by the strongly nonlinear character of (13). In the next section we try to approximate an appropriate sieve based on the original conception in [1] and examine the consequences of applying such in an originally homogeneous space like in the [0, 1] real interval. We are also try to elucidate what it would mean averaging over such spaces for wave like phenomena and their large scale averaging.

The fact that these numbers appear to be less than the reals already implies a kind of disconnected topology over the real line but in higher dimensional spaces it might point towards multiply connected domains possibly with a noisy, multifractal structure like a finely thin network. Some indications for this we shall present later on.

## 2 Self similarity in the entropic hierarchy

The hierarchical method of global maps is a "holistic" approach to certain computable properties and their asymptotic behavior for as long as it makes possible to derive analytic formulas from small initial segments otherwise it becomes computationally hard. In the case of the FIN set we see that the number of unique elements for the entropy function increases linearly for every odd-even pair of integer exponents $k$. One then needs to go to large sequences to check the nature of the FIN subset.

While this is possible with a direct execution of the diminishing probability protocol offered in [1] we can first examine the variability of the entropy global map. For each $S_k$ interval, with lowering entropy there are always two halves in $S_{k+1}$ that we may choose depending on the next bit being 0 or 1. Since all entropies are symmetric with respect to these two halves, one can only make a random choice for the subset of immediately lower value. This leads to making a direct comparison of the half of new values for any $H(S_{k+1}) - H(S_k)$ as shown in figure 3 for the transition from $S_8$ to $S_9$. A search over pairs of such intervals reveals a slow growth of an excess of negative values over previous positions in either of two halves corresponding to the addition of either a single "0" or "1" symbol. Assuming then, an algorithm capable of navigating between intervals by following a sequence of transitions it should be possible to locate those thin threads where the partial summands of successive entropy variations become stationary tending to a constant value. Instead, one can just isolate unique values for every $S_k$ such that the total entropy would obey a compositional law. We notice that the number of unique values increases linearly for every odd-even pair of exponents as a result of the digit-sum structure.

We then derive a scaling relation for the $h(x)$ function. Given an affine scaling as

$$v \to v' : k \to lk \vee \Sigma_2(v') \to l\Sigma_2(v) \quad (14)$$

for some $v \in S_k, v' \in S_{lk}$ it is trivial to deduce the scaling relation

$$h(l\Sigma_2(v)) = lh(\Sigma_2(v)) + (\Sigma_2(v))h(l) \quad (15)$$

For the complementary function $h(k - l\Sigma_2(v))$ entering $H^0$ we make use of the symmetry and since the new length is again $l(k - \Sigma_2(v))$ we still have $h(l\Sigma_2(2^k - v - 1)) = h(l\Sigma_2(\overline{v}))$ and the scaling relation remains the same for the complementary integer. From (13) and its complement we also deduce the analogous relation for $H^0$ taking into account that $(\Sigma_2(v) + \Sigma_2(\overline{v}))h(l) = kh(l) = kl\log_2(l)$ as

$$H^0(l\Sigma_2(v)) = lH^0(\Sigma_2(v)) + kl\log_2(l) \quad (16)$$

From the scaling relations we immediately obtain the new total entropy as

$$H_{lk}(v') = \log_2(lk) - \frac{1}{k}H^0(v) - \log_2(l) \quad (17)$$

Using (6) to replace $H^0(v)$ in (16) in terms of $H_k(v)$ leads to a fixed point condition

$$H_{lk}(v') = H_k(v) \quad (18)$$

We conclude that the set of entropies is self-affine across the hierarchy with each choice of $l$ spanning the hierarchy in subsets of equivalence classes. We notice that the relation holds also for fractional values of $l$ whenever $lk \in N$. Writing the fractional form explicitly as $l = \mu/\lambda$, $\mu, \lambda \in N$ leads to the condition that $\lambda$ must be a divisor of $k$. Since any pair of digit-sums from different string lengths can form a ratio with $\mu = \Sigma_2(v'), v' \in S_{lk}, \lambda = \Sigma_2(v), v \in S_k$, the fractional condition is equivalent to the demand that $k = n\Sigma_2(v)$. This is in accord with the need to take into account the generation of new values for entropies across the hierarchy together with the repetitiveness of the digit-sum function. The assumption of the existence of entropy bound must then lead to a particular set of equivalence classes. Moreover, from (18) we deduce that there is always a countable infinity of symbols available to satisfy the bound constraint.

**4 Discussion and Conclusions**

The really acute problem with a standard Cantorian continuum is its unverifiable nature by any finite memory device or entity with a finite memory capacity. In the simplest case, one might consider whether the question of the electron's charge being a rational number being meaningful in principle. As a matter of fact, such a continuum is that much large that one could conceive of the whole multiverse being encoded into a neighborhood of zero! Yet, one cannot really claim that one can just answer the previous as strictly possible or impossible, because of its non-empirical nature hence making such a question part of a set of undecidable propositions – as for instance, the proof for the existence of a God/s. This does not at the moment conflict with the Bekenstein bound argument since the practical problem is how to saturate such a bound in our everyday space if it exists, unless future experimental research proves otherwise.

One then is left with only a pair of options, the first being the already adopted one according to which a continuum serves as a best approximation for an elastic medium where wave and other equations of the most general type can most easily be studied given the two centuries development of appropriate mathematical armory. The other option is to attempt to find alternative descriptions based on the fact that other, sufficiently elaborate mathematical constructs might be able to imitate certain physical phenomena at least as an alternative interpolation scheme ("Black Box" problem) At this point one should be reminded that even a proposition that the physical universe as a whole is isomorphic to a "formal system" (and thus it contains a "lesser" symbolic model) is in itself a premise and it could or should be considered as a metaphysical proposition.

Philosophical ramifications aside, there is a critical question remaining after the previous analysis on whether the removal process mentioned in section 2, can still leave enough disks on a plane or in $R^3$ touching each other so as to keep at least a multiply connected topology, possibly with a filamentary structure which is reminiscent also of certain percolation models. Lastly, we notice the peculiar similarity of the last argument with recent astrophysical observations on strong filamentation in the long range macroscopic distribution of matter in the universe [25], [26]. Interestingly, some recent algorithms that appeared for the study of filamentation [27], [28], make use of the Morse-Smale complex[29] which could be adapted in an entropic potential to reach different levels of maximization for the bit length $k$ to reveal a fractal lattice in successive approximations across any bounded domain. One could at least conceive such as the complement of a fat fractal set of measure one, as the standard Smith-Volterra-Cantor set [30], [31]. It is then questionable whether large scale gravitational dynamics contains nature's attempt at concentrating around appropriate arithmetical structures in $R^3$ where there are more possibilities than a simple 1D example could reveal as is the case for in the Bing-Whitehead nonstandard, wild Cantor set [32], [33].

Further study of these subjects would have to combine advanced tools from metric number theory with so called Diophantine approximations [34], algebraic topology and other fields which are beyond the scope of this short report. Adopting Gisin's proposal then leads to an equally interesting question of such large scale structure being an indirect reflection of a similar substrate in space itself, in accord with the most recent understanding of limitations over physical space.

**References**

[1] N., Gisin, "Indeterminism in Physics, Classical Chaos and Bohmian Mechanics. Are Real Numbers Really Real?", David Bohm Centennial Symposium, London, Oct. 2017. arXiv:1803.06824 [quant-ph]

[2] T. E. Raptis, "Viral Turing Machines, computation from noise and combinatorial hierarchies.", Chaos, Sol. Fract., 104 (2017), 734-740.

[3] H. Trollope, "An explicit expression for binary digit sums.", *Math. Mag.*, 41, (1968) 21-25.

[4] H. Delange, "Sur la fonction sommatoire de la fonction Somme de Chiffres.", *Enseignemant Math.*, (2) 21 (1975) 31-47.

[5] P. Flajolet *et al.*, "Mellin transforms and asymptotics: Digital Sums.", *Theor. Comp. Sci.*, 123 (1994) 291-314.

[6] P. Flajolet *et al.*, "Mellin transforms and asymptotics: Harmonic Sums.", *Theor. Comp. Sci.*, 144 (1995) 3-58.

[7] A. O. Gelfond, "Sur les nombres qui ont des proprieties additives et multiplicatives donnees.", *Acta Arithmetica*, 13 (1968) 259-265.

[8] J. Besineau, "Independence statistique d' ensembles lies a la fonction 'sommes de chiffres'.", *Acta Arithmetica*, 20 (1972) 401-416.

[9] D.-H. Kim, "On the joint distribution of q-additive functions in residue classes.", *J. Num. Theory*, 74 (1999) 307-336.

[10] C. Mauduit, J. Rivat, "Le sommes de chiffres de carres.", *Acta Mathematica*, 203(1) (2009) 107-148.

[11] C. Mauduit, J. Rivat, "Sur un probleme de Gelfond: la somme de chiffres de nombres premieres." 171(3) (2010) 1591-1646.

[12] F. T. Adams-Watters, F, Ruskey, "Generating Functions for the Digital Sum and Other Digit Counting Sequences.", *J. Int. Seq.* 12 (2009) 09.5.6.

[13] R. Bellman, H. N. Shapiro. "On a problem in additive number theory.", *Annals of Mathematics*, 49(2) (1948) 333–340.

[14] C. Mauduit, A. Sarkozy, "On the Arithmetic Structure of Sets Characterized by Sum of Digit Properties.", *J. Num. Theory*, 61 (1993) 25-38.

[15] M. G. Madritch, "The sum-of-digits function of canonical number systems: Distribution in residue classes.", *J. Num. Theory*, 132 (2012) 2756-2772.

[16] Y. Kamiya, L. Murata, "Relations among arithmetic functions, automatic sequences, and sum of digits functions induced by certain Gray codes.", *J. Theor. Nombres*, 24 (2012) 307-337.

[17] P. Flajolet, L. Ramshaw, "A note on Gray code and odd-even merge.", *SIAM J. Comput.*, 9 (1980) 142-158.

[18] C. Müllner, "Exponential Sum Estimates and Fourier Analytic Methods for Digitally Based Dynamical Systems." PhD thesis, (2017) Technische Universität Wien and L' Université d' Aix-Marseille.

[19] https://oeis.org/A000120

[20] G. E. Revesz, (2009) "Lambda-calculus, Combinators and Functional Programming", Cammbridge Tracts in Theor. Comp. Sci.

[21] M. Kruppel, "Takagi's continuous nowhere differentiable function and binary digital sums." *Rostock. Math. Kolloq*, 63 (2008) 37-54.

[22] T. Takagi, "A Simple Example of the Continuous Function without Derivative", *Proc. Phys. Math. Japan*, 1 (1901) 176–177.

[23] P. Aalart, K. Kawamura, "The Takagi function: a survey.", *Real Anal. Exchange*, 37(1) (2011) 1-54, arXiv:1110:1691 [math.CA].

[24] J. C. Lagarias, "The Takagi Function and Its Properties." in "*Functions and Number Theory and Their Probabilistic Aspects*", K. Matsumoto, (2012) RIMS Kokyuroku Bessatsu, 153-189.

[25] A. P. Fairall *et al.*, "Visualization of Nearby Large-Scale Structures" in "*Unveiling large-scale structures behind the Milky Way*.", ASP Conf. Series, 67 (1994).

[26] I. Horvath *et al.,* (2013). "The largest structure of the Universe, defined by Gamma-Ray Bursts". *7th Huntsville Gamma-Ray Burst Symposium* (2013).

[27] T. Sousbie, "The persistent cosmic web and its filamentary structure – I. Theory and implementation.", and "The persistent cosmic web and its filamentary structure – II. Illustrations.", *Mon. Not. R. Astron. Soc.*, 414(1) (2011) 350-383 and 384-403. Program Hosting Website: http://www2.iap.fr/users/sousbie/web/html/indexd41d.html

[28] N. Shrivashankar, "Felix: A Topology based Framework for Visual Exploration of Cosmic Filaments.", *IEEE Transactions on Visualization & Computer Graphics*, 22(6) (2016) 1745-1759.

[29] M. Maller, "Fitted diffeomorphisms of non-simply connected manifolds" *Topology* , 19 (1980) 395–410.

[30] E. Ott, E. "Fat Fractals." §3.9 in "*Chaos in Dynamical Systems.*" (1993) NY, Cambridge Univ. Press, 97-100.

[31] M. Balcerzak, A. Kharazishvili, "On uncountable unions and intersections of measurable sets", *Georgian Math. J.*, 6 (3) (1999) 201–212.

[32] D. G. Wright, "Bing-Whitehead Cantor Sets.", *Fundamenta Mathematicae,* 132(2) (1987) 105-116.

[33] D. Garity *et al.*, "Distinguishing Bing-Whitehead Cantor Sets.", *Trans. Amer. Math. Soc.* 363(2) (2011) 1007-1022.

[34] S. Lang, (1995) "*Introduction to Diophantine Approximations.*", NY, Springer-Verlag.

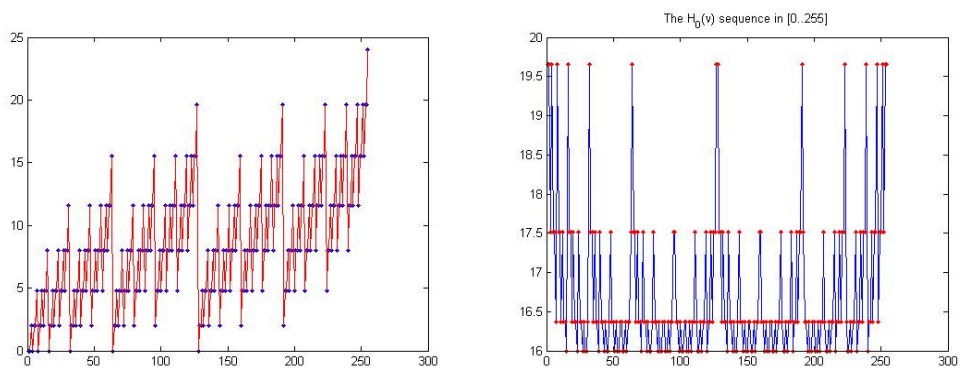

**Fig. 1** (a) The increasing entropic term of eq. (4), (b) the symmetrized fractal sequence $H_0$ on the same interval.

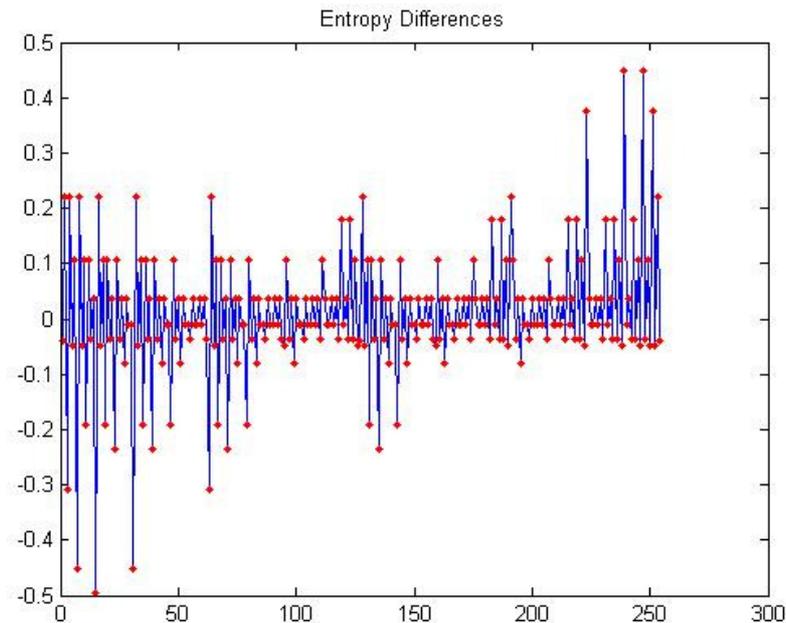

**Fig. 2** Differences in variation of entropy global maps in a single symbol transition between the $S_8$ and the first half of $S_9$ intervals.